\documentclass[a4paper,11pt]{article}
\pdfoutput=1 

\usepackage{jheppub} 

\usepackage[T1]{fontenc} 
\usepackage{subfigure}

\title{\boldmath Vector Boson Fusion Topology and Simplified Models for Dark Matter Searches at Colliders}

\author{Santiago Duque-Escobar,}
\author{Daniel Ocampo-Henao,}
\author{José D. Ruiz-Álvarez}\author{A. Uthor}
\affiliation{Instituto de Física, Universidad de Antioquia,\\ A.A. 1226, Medellín, Colombia}

\emailAdd{asantiago.duque@udea.edu.co}
\emailAdd{daniel.ocampoh@udea.edu.co}
\emailAdd{josed.ruiz@udea.edu.co}

\abstract{In this paper, we study the possible searches at colliders using Vector Boson Fusion topology in the context of Simplified Models signatures. We examine the possible physics reach of these searches with regard to monojet-type searches, and determine how these two signatures are complementary. We determine the generic characteristics for dark matter signatures in the LHC if the underlying physics imply Vector Boson Fusion type of production.}

\begin{document} 
\maketitle
\flushbottom

\section{\label{sec:level1} Introduction}

One of the most striking challenges in contemporary physics is to determine the nature of around a quarter of the matter content of the universe. This fraction is usually referred to as dark matter. We have several sources of evidence pointing to the existence of an unknown form of matter from different types of astronomical measurements \cite{Griest:1993cv, rubin1983rotation, minchin2005dark, Zacek:2007mi}. However, direct and indirect searches have failed to find a particle which could explain the dark matter abundance. 

The quest towards dark matter discovery can be also pursued in particle colliders as the Large Hadron Collider (LHC). Several searches have been done by ATLAS and CMS experiments \cite{CMS:2017zts,CMS:2019san, CMS:2019ykj, ATLAS:2021kxv, ATLAS:2020yzc, CMS:2020ulv}, but still no positive indication of the production of a dark matter particle in the LHC has been found. 

Several theoretical frameworks have been developed to explore dark matter searches at colliders. More specifically, the simplified models scheme \cite{abdallah2014simplified, Abdallah:2015ter} constitute a useful benchmark modelling of dark matter production at the LHC that have been introduced in recent years. The simplified models provide a rather free theoretical approach and simple interpretation framework. Therefore, it is interesting to envisage all possible signatures for dark matter searches within the simplified model approach. It is very important to identify all the possibilities that can be pursued at the LHC within this modelling.

We study the possible Vector Boson Fusion (VBF) signatures that can be achieved within simplified models with dark matter production. We optimize the selections to enhance signal separation in VBF signatures with regard to standard model backgrounds. Our findings are then compared to monojet-type searches and show that VBF signatures are complimentary in dark matter searches.

In this paper we first discuss the simplified models and possible VBF signatures in section~\ref{sec:level2}, then we show the VBF cuts developed in section~\ref{sec:level3}. In section~\ref{sec:level4}, we present our results and discuss them in regard of monojet-type searches, and we conclude in section~\ref{sec:level5}.

\section{\label{sec:level2} Simplified models for dark matter searches at the LHC}

Within the scope of performing searches for dark matter in colliders, many different approaches have been proposed. Possibly the first approach considered has been the neutralino candidate within a MSSM theoretical framework \cite{Csaki:1996ks, Haber:1997if}. However it has been quickly identified that such an approach is limited and present strong theoretical restrictions. Many more generic models have been proposed as the Inert Dark Matter model \cite{Dolle:2009fn, Goudelis:2013uca}. These models too are limited for searches within the LHC as they are formerly constrained from a theory perspective. 

Finally, the last step before the arrival of simplified models has been the effective field theory (EFT) approach \cite{georgi1993effective, Burgess:2007pt}. In this theory, the mediator that communicates the dark matter candidate with the SM is very heavy, and thus escaping the energy reach of the LHC. This has strong implications for the kinematic properties of the dark matter production signatures, making it extremely difficult to search for a dark matter candidate with a rather light mediator.

At this point, the simplified models were proposed as an intermediate solution among the EFT and complete models. The simplified models allow having a quite complete theoretical picture with a very unrestricted modelling and an easy framework for reinterpretations. Therefore, the simplified models can be used to inspire and design searches at the LHC, but can also be used to interpret LHC results and also export them into the context of complete models, as the MSSM.

The simplest formulation is a simplified model with a fermionic dark matter candidate and a scalar or vector mediator. The mediator couples only to quarks and dark matter. In the first case to allow the production of the additional mediator in the LHC from proton-proton collision. The second coupling ensures the decay of the mediator into dark matter particles. The corresponding Lagrangians, following definitions from \cite{Abdallah:2015ter, Backovic:2015soa, Neubert:2015fka}, are given in equations \ref{fig:lag11}-\ref{fig:lag13} for the scalar mediator case and \ref{fig:lag21}-\ref{fig:lag23} for the vector mediator case.

\begin{align}
    \mathcal{L}_{DM}^{Y_{0}} &= g_{\chi}\bar{\chi}\chi Y_{0} \label{fig:lag11} \\ 
    \mathcal{L}_{QCD}^{Y_{0}} &= \frac{g_{q}}{\sqrt{2}}\sum_{i,j}(y_{ij}^{u}\bar{u}_{i}u_{j}+y_{ij}^{d}\bar{d}_{i}d_{j})Y_{0} \label{fig:lag12} \\ 
    \mathcal{L}_{h}^{Y_{0}} &= m_{Y_{0}}g_{S1}|\phi|^{2}Y_{0}+g_{S2}|\phi|^{2}Y_{0}^{2}
    \label{fig:lag13} \\
    \mathcal{L}_{EW}^{Y_{0}} &= \frac{1}{\Lambda}m_{Y_{0}}[g_{h3}^{S}(D^{\mu}\phi)^{\dagger}(D_{\mu}\phi) +g_{B}^{S}B_{\mu\nu}B^{\mu\nu}+g_{W}^{S}W_{\mu\nu}^{i}W^{i,\mu\nu}]Y_{0}
    \label{fig:lag14}
\\
    \mathcal{L}_{DM}^{Y_{1}} &= g_{\chi}\bar{\chi}\gamma_{\mu}\chi Y_{1}^{\mu} \label{fig:lag21} \\ 
    \mathcal{L}_{QCD}^{Y_{1}} &= \sum_{i,j}(g_{ij}^{u}\bar{u}_{i}\gamma_{\mu}u_{j}+g_{ij}^{d}\bar{d}_{i}\gamma_{\mu}d_{j})Y_{1}^{\mu} \label{fig:lag22} \\ 
    \mathcal{L}_{EW}^{Y_{1}} &= g_{V}\frac{i}{2}(\phi^{\dagger}(D_{\mu}\phi)-(D_{\mu}\phi)^{\dagger}\phi) Y_{1}^{\mu}
    \label{fig:lag23}
\end{align}

In equations \ref{fig:lag11} and \ref{fig:lag21}, $g_{\chi}$ denotes the coupling between the dark matter particle candidate $\chi$ and the mediator $Y_{0}$ or $Y_{1}$. On the other hand, in equation \ref{fig:lag12}, $g_{q}$ denotes a generic coupling between the new mediator and the up and down type quarks from the SM; while in equation \ref{fig:lag22}, $g_{ij}^{u}$ and $g_{ij}^{d}$ are the quarks couplings to the vector mediator to up and donw-type quarks correspondingly. Equation \ref{fig:lag13} shows the linear and quadratic couplings of $Y_{0}$ with the Higgs scalar field, correspondingly $g_{S1}$ and $g_{S2}$. In addition, $m_{Y_{0}}$ corresponds to the mass of the scalar mediator. The interactions between the scalar mediator and the SM bosons are included through effective dimension 5 operators described in equation \ref{fig:lag14}. $g_{h3}^{S}$ denotes an additional coupling of the scalar mediator with the Higgs field, while $g_{B}^{S}$ and $g_{W}^{S}$ denote the couplings with SM bosons. The parameter $\Lambda$ is set to 10 TeV as some large energy scale. Finally, in equation \ref{fig:lag23}, $g_{V}$ denotes the coupling of the vector mediator $Y_{1}$ with the SM Higgs scalar field and the electroweak bosons. The $\phi$ field corresponds to the Higgs  doublet. The equation \ref{fig:lag13} gives rise to mixings between the Z boson and the vector mediator. It should be kept in mind that we have taken a simplified version of the Lagrangian for our studies, but in a generic approach pseudoscalar and axial-vector couplings could be also written.

\begin{figure}[h]
    \begin{center}   
    \subfigure[]
    {
        \includegraphics[width=0.45\columnwidth]{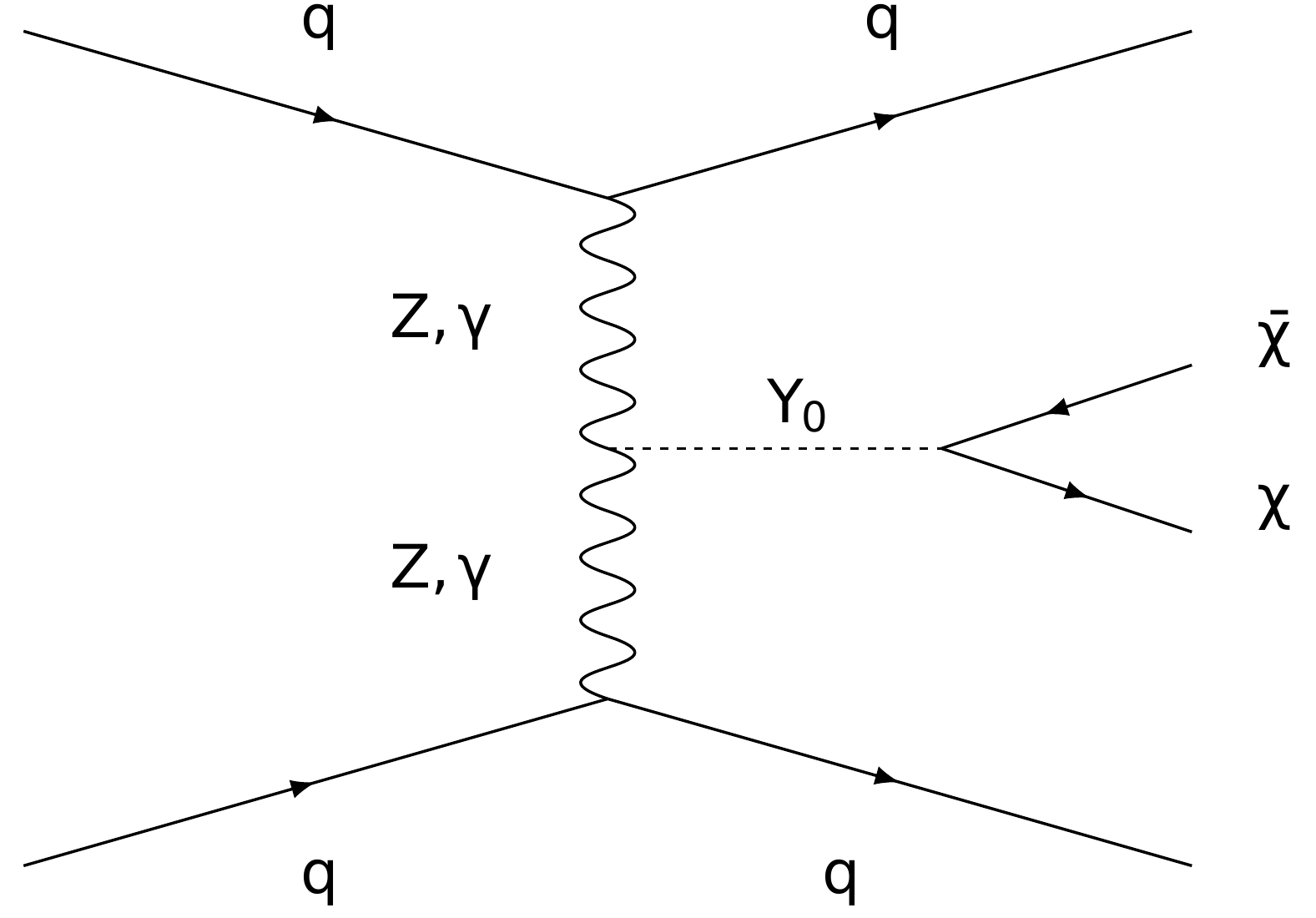}
        \label{fig:feynman_a}
    }
    \subfigure[]
    {
        \includegraphics[width=0.45\columnwidth]{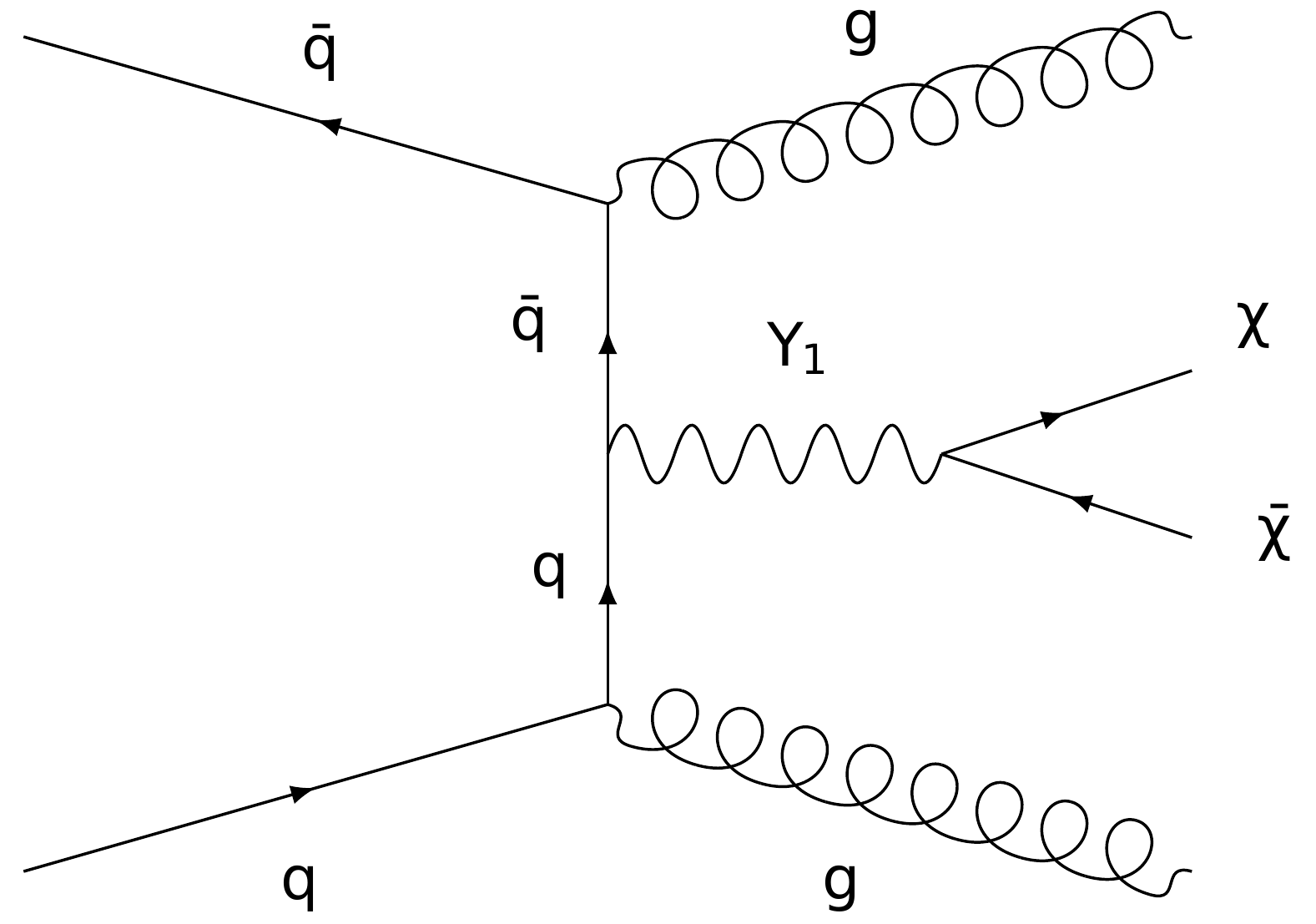}
        \label{fig:feynman_b}
    }
    \caption{\label{fig:feynman}Main processes for dark matter production in a VBF topology: a) Model with a scalar mediator. b) Model with a vector mediator.}
    \end{center}
\end{figure}

The main processes for dark matter production in a VBF topology for each of these models are shown in Fig.~\ref{fig:feynman}.

The two diagrams share a common structure, but for the scalar mediator case the two final hadronic particles are quarks while in the vector mediator case are gluons. Additionally, in the scalar mediator case the particles mediating the process are vector bosons from the SM and in the vector mediator case are quarks, which of course also implies a coupling of the new mediator to standard model bosons. Therefore, the scalar mediator case is an authentic VBF process, very similar to the VBF Higgs production mechanism. This fact allows us to explore the VBF production of dark matter in a Higgs-like mechanism, but with mediator masses that can be much greater. The vector mediator case is more exotic to dark matter searches, letting us explore new possible scenarios.

Moreover, the simplified models allow us to explore a very wide range of mediator masses as well as mediator couplings. These factors are key to our study and will be fully detailed in the next sections. We would like to stress that we use the simplified models approach as a tool to study the kinematics of VBF topology with different mediators spin and mass in the context of the LHC. Therefore, we don't examine the possible constraints of the models from other observables.

\section{\label{sec:level3} Vector Boson Fusion searches}

Since one of the main features of dark matter is its extremely small interaction with normal matter, all searches at colliders rely on the production of some visible particle recoiling against the dark matter, which will subsequently be tagged in the detector as missing transverse momentum or energy $p_T^{miss}$. A common strategy to look for such events is the monot-jet search, already well studied in the context of collider searches~\cite{CMS:2017zts}. It is also important to remark that the simplified models approach has born in the context of mono-jet searches.

VBF events are characterized by two interacting vector bosons and two deflected partons from the initial partons resulting in two forward jets plus $p_T^{miss}$. The main characteristics of the VBF topology that have been identified in the literature are a high invariant mass for the two VBF jets, a geometric location of these two jets in opposite hemispheres of the detector and a high $\eta$ separation. This topology is an alternative and complementary strategy to look for dark matter at colliders. The VBF topology allows lower kinematic thresholds, resulting in non-$p_T^{miss}$ based event selection~\cite{CMS:2018yfx}. The utility of the VBF topology has been shown in other beyond standard model contexts such as Higgs Portal dark matter and MSSM~\cite{Dutta:2017lny, CMS:2019san}.

\subsection{Simulated event samples}

The signal has been simulated from proton-proton collisions at $\sqrt{s} = 13 \; \text{TeV}$ in a VBF topology with scalar or vectorial mediators. The signal samples were produced for several combinations of dark matter and mediator masses points ($m_\chi$, $m_Y$), ranging from 10 to 1000 GeV and from 100 to 5000 GeV, respectively. The signal model used is a simplified dark matter model from {\textsc{FeynRules} model database}~\cite{Abdallah:2014hon,Buckley:2014fba,Abdallah:2015ter}. All the models were used in their UFO~\cite{Degrande:2011ua} implementation.

Electroweak bosons produced in association with jets are the primary backgrounds for this search, particularly, the production of a $Z$ boson and jets, followed by the decay of the $Z$ boson into neutrinos.  In the search for invisible decays of the Higgs boson~\cite{CMS:2018yfx} is found that the main background contribution is coming from $Z$ and $W$ bosons production, while other standard model backgrounds only contribute with 3\% of the total background composition. In the present study we have simulated only Drell-Yan plus jets (up to 2) background and following~\cite{Dutta:2017lny} reasoning and teh results from~\cite{CMS:2018yfx}, we consider $W+$jets background kinematics are similar to Drell-Yan, and that the final background contribution is composed 70\% of Drell-Yan events and 30\% of $W+$jets.

The partonic processes for signal and background events have been produced with {\textsc{MadGraph5\_aMC}} (v2.8.2)~\cite{Alwall:2014hca}. The showering and hadronization have been performed by {\textsc{PYTHIA8}}~\cite{Sjostrand:2006za,Sjostrand:2014zea}, and the detector simulation has been done by {\textsc{Delphes}} (v3.4.2)~\cite{deFavereau:2013fsa}. We have used the same packages to derive a k-factor for Drell-Yan plus jets background to take into account NLO effects on the cross-section prediction. We found a k-factor of 1.2 and applied it accordingly to our results.

\subsection{Event selection}

We have developed an event selection optimizing the significance, defined as $\frac{S}{\sqrt{S+B}}$, using signal events both from scalar and vectorial mediators signals and from Drell-Yan background. We have noted that for both signals considered the optimization had the same results and therefore we end up with one selection regardless of the signal mediator type. The selection begins with very basic criteria mainly driven by well known detector requirements. This baseline selection is described in Table~\ref{tab:baselineselection}. The total hadronic energy is defined as $H_{T}=\sum p_{T} (j)$ which is the scalar sum of the transverse momentum of all the jets in the event which have a $p_{T}>30$~GeV. The criterion of $H_{T}>200$~GeV is imposed in order to assure a minimum quantity of energy to assure signal events passing the detector trigger. This requirement prevents us to rely on missing transverse momentum for the trigger. We have seen that signal events don't have very high missing transverse momentum, and therefore the $H_{T}$ criterion is very useful to avoid losing a large amount of signal events by the trigger.

\begin{table}[tbp]
\centering
\begin{tabular}{|l|l|}
\hline
\textbf{Criterion}        &  \\ \hline
Number of jets      & \textgreater{} 1 \\

$\eta(j_1) \cdot \eta(j_2)$  &  \textless{}  0 \\
Leading jets $p_T$                                                             & \textgreater{} 30 GeV      \\
Leading jets $|\eta|$                                                             & \textless{} 5  \\

$H_T$                                             & \textgreater{} 200 GeV          \\
$p_{T}^{miss}$                                             & \textgreater{} 50 GeV          \\ \hline
\end{tabular}
\caption{\label{tab:baselineselection} Baseline selection}
\end{table}

\begin{figure}[htbp]
    \subfigure[ ]
    {
        \includegraphics[width=0.5\columnwidth]{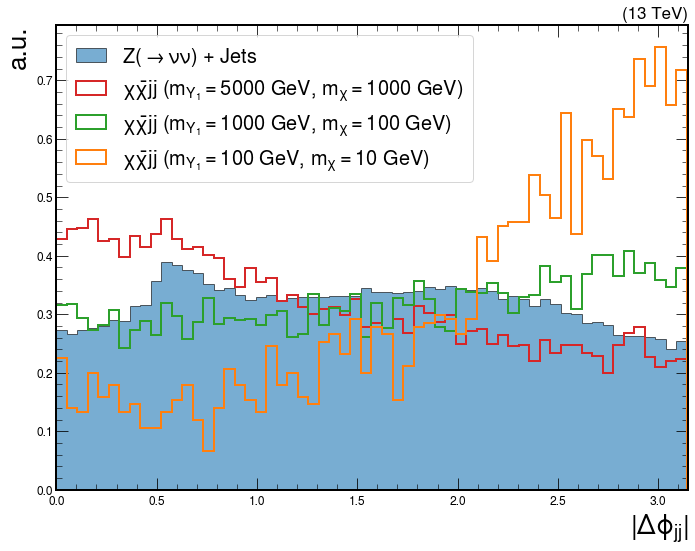}
        \label{fig:a}
    }
    \subfigure[ ]
    {
        \includegraphics[width=0.5\columnwidth]{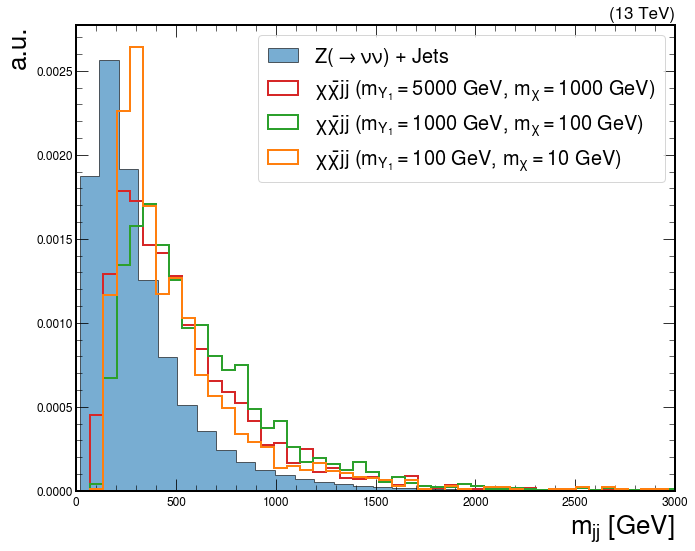}
        \label{fig:b}
    }
    \caption{\label{fig:deltaphimjj} Signals for vectorial mediator and background events distribution for [a] $|\Delta \phi_{jj}|$ variable on events passing the baseline selection, and [b] $m_{jj}$ for events passing the baseline selection and $|\Delta \phi_{jj}|>2.3$}
\end{figure}

For events surviving the baseline selection, we have studied several variables that could potentially increase the signal over background separation. The most discriminant variables are found to be the azimuthal angle difference among the two leading jets $\Delta \phi_{jj}$, their invariant mass $m_{jj}$ and their pseudorapidity separation $\Delta \eta_{jj}$. The $\Delta \phi_{jj}$ show some dependence on the mediator mass while $m_{jj}$ variable has a homogeneous behavior for different mediator masses. Regarding the classical definition of VBF topology properties the mos important finding is the dependence on the mediator masses of the pseudorapidity separation of the two leading jets. In consequence, we start our main selection based on criteria on the azimuthal difference of the leading jets and their invariant mass. Figure~\ref{fig:deltaphimjj} show the distribution of signal and background events for $|\Delta \phi_{jj}|$ after baseline selection is applied and for $m_{jj}$ after applying a selection also on the azimuthal angle separation.

\sloppy From these two first requirements we apply a control criterion intended not for separation of signal with regard to electroweak backgrounds but to actually control the possible contribution from QCD processes to the background. The applied criterion is ${min|\Delta \phi(p_T^{miss}, j_i)|>0.5}$ for the first four leading jets, $i=\{1,2,3,4\}$. This cut ensures having real missing transverse momentum and thus rejecting efficiently QCD events \cite{CMS:2019san}, \cite{CMS:2018yfx}.

Finally, we have noted that the variable $\Delta \eta_{jj}$ greatly depends on the mass of the mediator. For high mediator masses, above 1 TeV, $\Delta \eta_{jj}$ is restricted to low values, while for low mediator masses, below 1 TeV, the $\Delta \eta_{jj}$ is much higher. The most known scenario for VBF topology is for Higgs SM production, which corresponds to our low mass mediator case, from which a high $\Delta \eta_{jj}$ value has been identified as a characteristic feature of VBF, however we see that the pseudorapidity difference for the two VBF jets is not fixed and actually highly anti-correlated to the mediator mass. This finding is also supported by the work done in \cite{Florez:2019tqr}.

\begin{figure}[b]
\includegraphics[width=0.5\columnwidth]{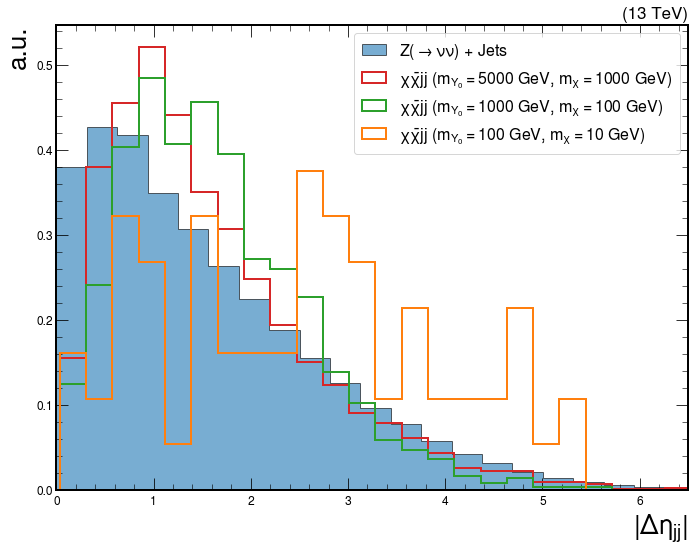}
\includegraphics[width=0.5\columnwidth]{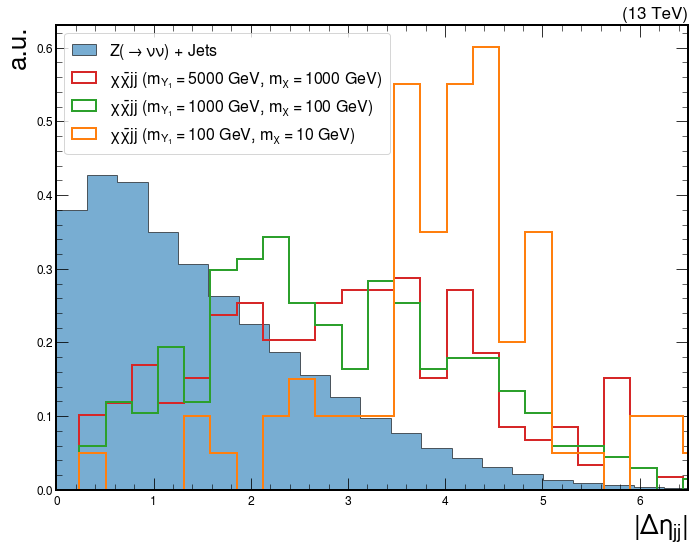}
\caption{\label{fig:deltaeta} Signals and background events distribution for $|\Delta \eta_{jj}|$ variable on events passing all the other criteria.}
\end{figure}

In correspondence with this finding, we define two analysis bins optimized to keep events for the two mediator mass scenarios described. Table~\ref{tab:cutselection} summarizes the event selection and Figure~\ref{fig:deltaeta} shows $| \Delta \eta_{jj}|$ for signal events and backgrounds after all other cuts.

\begin{table}[tbp]
\centering
\begin{tabular}{|l|ll|}
\hline
\textbf{Feature}              & \multicolumn{2}{l|}{\textbf{Value}} \\ \hline 
$|\Delta \phi_{jj}|$    & \multicolumn{2}{l|}{\textgreater{} 2.3}          \\
$m_{jj}$         & \multicolumn{2}{l|}{\textgreater{} 1000 GeV} \\ 
min$|\Delta \phi(p_T^{miss}, j_i)|$ & \multicolumn{2}{l|}{\textgreater{} 0.5}  \\ 
$| \Delta \eta_{jj}|$ & \textless{} 2.5 & or \textgreater{} 2.5          \\
\hline   
\end{tabular}
\caption{\label{tab:cutselection} Event selection. The last line specifies the two bin selections optimized for high and low mediator masses.}
\end{table}

\section{\label{sec:level4} Results and discussion}

From the selection developed we establish its reach considering different dark matter and mediator masses. The simplified models approach only depends on five parameters, the dark matter mass, the mediator mass, and the couplings of the mediator to quarks, to SM bosons and dark matter. We first establish the significance of the selection for fixed couplings for some benchmark masses. We then analyze the reach of the analysis for fixed masses while variating the couplings.

\begin{figure}[h]
\centering
\includegraphics[width=0.6\columnwidth]{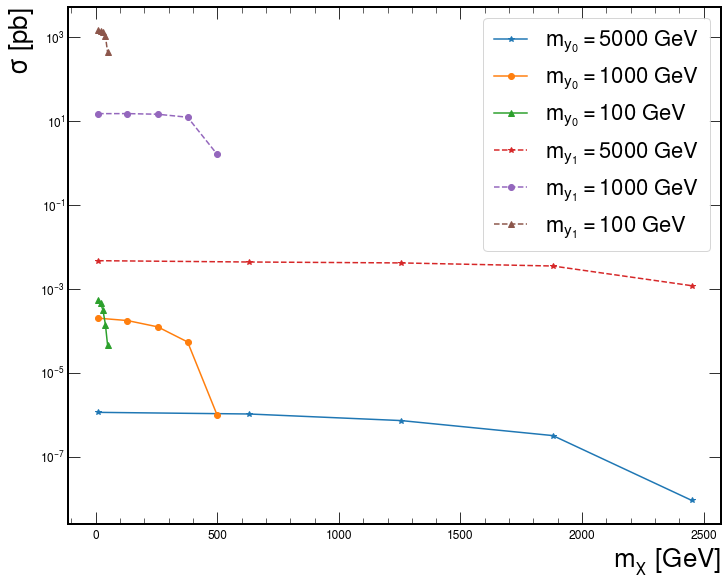}
\caption{\label{fig:xsections} Signals cross-sections for fixed mediator mass, as described in the legend, setting $g_{\chi}=1$, $g_{V}=1$ and $g_{q}=0.25$ as a function of the dark matter candidate mass.}
\end{figure}

In Tables~\ref{tab:cutflow},~\ref{tab:significance} and~\ref{tab:efficiency} are shown the number of expected events, the significance and cut efficiency as the selection criteria are applied for the three benchmark masses for signal and background for $150\; \text{fb}^{-1}$. The expected number of events are calculated using the cross-sections obtained from MadGraph setting couplings of the mediator to dark matter $g_{\chi}=1$, to standard model bosons $g_{V}=1$ and to quarks $g_{q}=0.25$. The cross-sections are shown for fixed mediator mass, fixed couplings and as a function of the dark matter mass in Figure~\ref{fig:xsections}. All cross-sections and event generation have been done at leading order level.

\begin{table*}
\centering
\resizebox{\columnwidth}{!}{%
\begin{tabular}{|l|c|c|c|c|c|c|c|}
\hline
 & \multicolumn{3}{c|}{Scalar $(m_{Y_{0}},m_{X})$ GeV} & \multicolumn{3}{c|}{Vector $(m_{Y_{1}},m_{X})$ GeV} & \\ \hline
\textbf{Selection}              & (100,10) & (1000,100) & (5000,1000) & (100,10) & (1000,100) & (5000,1000) & B \\ \hline
Baseline    & 32.84	 & 76.07 & 1.34 & 1.16e+7 & 4.67e+5 & 170.80 & 6.48e+8         \\
$|\Delta \phi_{jj}|$    & 29.19 & 66.76 & 1.33 & 5.89e+6 & 1.43e+5 & 37.47 & 1.79e+8          \\
$m_{jj}$      & 14.23 & 52.05 & 1.33 & 5.07e+5 & 1.89e+4 & 4.65 & 2.60e+7  \\ 
min$|\Delta \phi(p_T^{miss}, j_i)|$   & 1.56 & 10.02 & 0.26 & 3.00e+5 & 1.02e+4 & 2.65 & 1.72e+7  \\ 
$| \Delta \eta_{jj}|$   & 0.72, 0.84  & 8.30, 1.72 & 0.22, 0.04& 2.43e+4, 2.76e+5 & 4.54e+3, 5.70e+3 & 1.07, 1.57 & 3.32e+6, 1.38e+7 \\ 
\hline
\end{tabular}
}
\caption{\label{tab:cutflow} Events after each selection applied for background and some signals, assuming a luminosity of $150\;\text{fb}^{-1}$.}
\end{table*}

\begin{table*}
\centering
\resizebox{\columnwidth}{!}{%
\begin{tabular}{|l|c|c|c|c|c|c|}
\hline
 & \multicolumn{3}{c|}{Scalar $(m_{Y_{0}},m_{X})$ GeV} & \multicolumn{3}{c|}{Vector $(m_{Y_{1}},m_{X})$ GeV} \\ \hline
\textbf{Selection}                      & (100,10)          & (1000,100)        & (5000,1000)       & (100,10)          & (1000,100)        & (5000,1000) \\ \hline
Baseline    							& 1.29e-3			& 2.98e-3			& 5.26e-5			& 4.51e+2			& 1.83e+1			& 6.70e-3	\\
$|\Delta \phi_{jj}|$    				& 2.18e-3			& 4.98e-3			& 9.94e-5			& 4.33e+2			& 1.06e+1			& 2.80e-3  \\
$m_{jj}$      							& 2.79e-3			& 1.02e-2			& 2.60e-4			& 9.84e+1			& 3.70e+0			& 9.11e-4  \\
min$|\Delta \phi(p_T^{miss}, j_i)|$   	& 3.76e-4			& 2.41e-3			& 6.26e-5			& 7.17e+1			& 2.45e+0			& 6.38e-4  \\
$| \Delta \eta_{jj}|$   				& 3.95e-4, 2.26e-4	& 4.55e-3, 4.63e-4	& 1.20e-4, 1.07e-5	& 1.32e+1, 7.35e+1	& 2.48e+0, 1.53e+0	& 5.87e-4, 4.22e-4  \\ \hline
\end{tabular}
}
\caption{\label{tab:significance} Significance defined as $\frac{S}{\sqrt{S+B}}$ for each signal and including the scale factor for background to include W+jets contribution, assuming a luminosity of $150\;\text{fb}^{-1}$.}
\end{table*}

\begin{table*}
\centering
\resizebox{\columnwidth}{!}{%
\begin{tabular}{|l|c|c|c|c|c|c|c|}
\hline
 & \multicolumn{3}{c|}{Scalar $(m_{Y_{0}},m_{X})$ GeV} & \multicolumn{3}{c|}{Vector $(m_{Y_{1}},m_{X})$ GeV} & \\ \hline
\textbf{Selection}              & (100,10) & (1000,100) & (5000,1000) & (100,10) & (1000,100) & (5000,1000) & B \\ \hline
Baseline    & 0.041 & 0.203 & 0.659 & 0.071 & 0.226 & 0.271 & 0.293       \\
$|\Delta \phi_{jj}|$    & 0.889 & 0.878 & 0.991 & 0.505 & 0.308 & 0.219 & 0.276          \\
$m_{jj}$      & 0.488 & 0.780 & 1.000 & 0.086 & 0.132 & 0.124 & 0.146  \\ 
min$|\Delta \phi(p_T^{miss}, j_i)|$   & 0.110 & 0.193 & 0.198 & 0.592 & 0.541 & 0.569 & 0.659  \\ 
$| \Delta \eta_{jj}|$   & 0.464, 0.536  & 0.828, 0.172 & 0.832, 0.168 & 0.081, 0.919 & 0.444, 0.556 & 0.405, 0.595 & 0.193, 0.807 \\ \hline
\end{tabular}
}
\caption{\label{tab:efficiency} Criteria efficiency for each signal point considered and for background.}
\end{table*}

\begin{figure}[h]
\includegraphics[width=0.5\columnwidth]{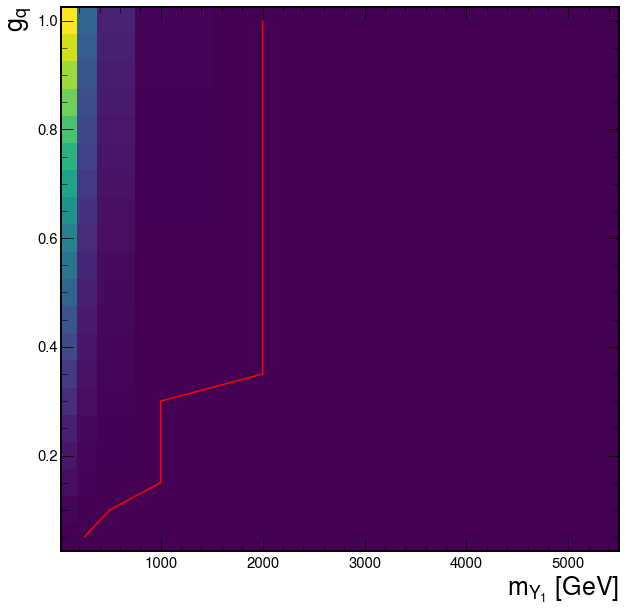}
\includegraphics[width=0.5\columnwidth]{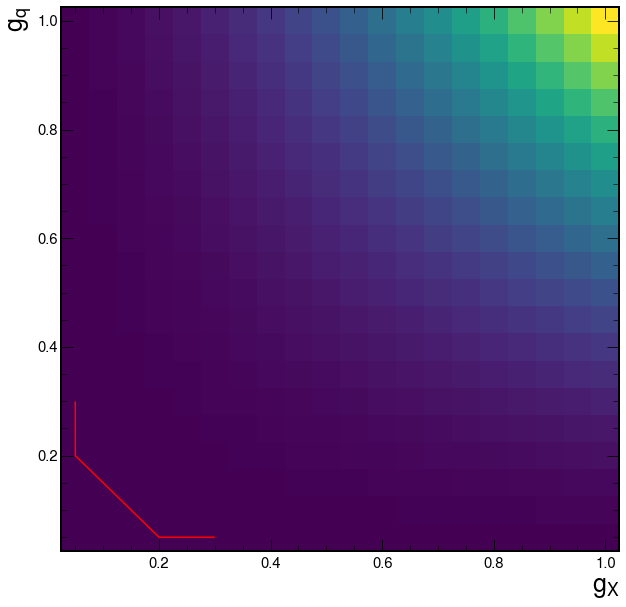}
\caption{\label{fig:exclusion} Exclusion reach of the selections for [left] $g_{q}$-mediator mass where the excluded region is from the red line for greater couplings and smaller mediator masses, and for [right] $g_{q}$-$g_{\chi}$ plane with the excluded region from the red line towards higher couplings.}
\end{figure}

Taking into account the dependence of the cross-sections to the couplings and to the mediator and dark matter mass, we have scanned the models parameters in order to identify the exclusion reach of the proposed selections. Figure \ref{fig:exclusion} show the excluded areas in the quark coupling-mediator mass plane while having fixed dark matter mass and coupling fixed to 10 GeV and 1.0 respectively. It also shows the exclusion in the quark and dark matter coupling of the mediator, with the mediator mass fixed to 1 TeV and dark matter mass fixed to 10 GeV. The excluded regions are built using a sensitivity $\frac{S}{\sqrt{S+B}}$ of at least 2, and are depicted in the plots as a red line. We only reach exclusion for the vectorial mediator scenario. The sensitivity is calculated using the total number of events for each signal sample and backgrounds after selection.

\section{\label{sec:level5} Conclusions}

From the results shown, we can state that VBF searches for dark matter show a reach up to 2 TeV in the mediator mass and down to 0.2 for the couplings in the vectorial mediator case. These results are compatible with the results achieved by monojet searches and complimentary because they impose limits on a different signature, which should be in nature if monojet process is allowed.

We find exclusion for the vector mediator case, however, and due to the small cross-sections, the exclusion is not achieved in the scalar mediator case. This might be improved in other theoretical and experimental scenarios. For example, a coupling of the scalar mediator to the SM W bosons can be added to the model, which would increase the signal production cross-section. On the experimental side, the High Luminosity LHC would give an improved statistical probability of producing signal events. Anyhow, an important finding is that for the simplified models the VBF signature does not differ strongly depending on the spin of the mediator. We should keep in mind that the scalar mediator mass can couple to the $W$ boson, in which case we can increase the production cross section.

In addition, we could also cover the parameter space of the model in the mediator and dark matter mass plane up to values closer to the diagonal with the dark matter mass equal to half of the mediator mass than the ones already covered by monojet searches. But this would also imply an assumption over the width of the mediator in order to assure stable cross-sections and efficiencies up to the diagonal. In other words, the reach of VBF search is better in the mediator and dark matter mass plane for narrow width mediators (less than 10 GeV) with regard to monojet searches.

Finally, we have found that in order to have sensitive VBF searches we might need to optimize the selections depending on the mediator masses explored due to kinematic differences as the ones we have found for the difference in the pseudorapity of the two VBF jets in the events.

\acknowledgments

The authors gratefully acknowledge the support of the Colombian Science Ministry \-Min\-Cien\-cias and Sos\-te\-ni\-bi\-li\-dad-UdeA. J.D.R.A also acknowledge the welcoming of the Instituto de F\'{i}sica Corpuscular (IFIC) from Valencia, Spain, during the completion of this work.


\bibliographystyle{JHEP}
\bibliography{paper.bib}

\providecommand{\noopsort}[1]{}\providecommand{\singleletter}[1]{#1}%

\providecommand{\href}[2]{#2}\begingroup\raggedright\begin{thebibliography}{10}

\bibitem{Griest:1993cv}
K.~Griest, \emph{{The Search for dark matter: WIMPs and MACHOs}},
  \href{https://doi.org/10.1111/j.1749-6632.1993.tb43912.x}{\emph{Annals N. Y.
  Acad. Sci.} {\bfseries 688} (1993) 390}
  [\href{https://arxiv.org/abs/hep-ph/9303253}{{\ttfamily hep-ph/9303253}}].

\bibitem{rubin1983rotation}
V.C.~Rubin, \emph{The rotation of spiral galaxies}, {\emph{Science} {\bfseries
  220} (1983) 1339}.

\bibitem{minchin2005dark}
R.~Minchin, J.~Davies, M.~Disney, P.~Boyce, D.~Garcia, C.~Jordan et~al.,
  \emph{A dark hydrogen cloud in the virgo cluster}, {\emph{The Astrophysical
  Journal Letters} {\bfseries 622} (2005) L21}.

\bibitem{Zacek:2007mi}
V.~Zacek, \emph{{Dark Matter}},  in \emph{{22nd Lake Louise Winter Institute:
  Fundamental Interactions}}, pp.~170--206, 2007,
  \href{https://doi.org/10.1142/9789812776105_0007}{DOI}
  [\href{https://arxiv.org/abs/0707.0472}{{\ttfamily 0707.0472}}].

\bibitem{CMS:2017zts}
{\scshape CMS} collaboration, \emph{{Search for new physics in final states
  with an energetic jet or a hadronically decaying $W$ or $Z$ boson and
  transverse momentum imbalance at $\sqrt{s}=13\text{ }\text{ }\mathrm{TeV}$}},
  \href{https://doi.org/10.1103/PhysRevD.97.092005}{\emph{Phys. Rev. D}
  {\bfseries 97} (2018) 092005}
  [\href{https://arxiv.org/abs/1712.02345}{{\ttfamily 1712.02345}}].

\bibitem{CMS:2019san}
{\scshape CMS} collaboration, \emph{{Search for supersymmetry with a compressed
  mass spectrum in the vector boson fusion topology with 1-lepton and 0-lepton
  final states in proton-proton collisions at $\sqrt{s}=$ 13 TeV}},
  \href{https://doi.org/10.1007/JHEP08(2019)150}{\emph{JHEP} {\bfseries 08}
  (2019) 150} [\href{https://arxiv.org/abs/1905.13059}{{\ttfamily
  1905.13059}}].

\bibitem{CMS:2019ykj}
{\scshape CMS} collaboration, \emph{{Search for dark matter particles produced
  in association with a Higgs boson in proton-proton collisions at $
  \sqrt{\mathrm{s}} $ = 13 TeV}},
  \href{https://doi.org/10.1007/JHEP03(2020)025}{\emph{JHEP} {\bfseries 03}
  (2020) 025} [\href{https://arxiv.org/abs/1908.01713}{{\ttfamily
  1908.01713}}].

\bibitem{ATLAS:2021kxv}
{\scshape ATLAS} collaboration, \emph{{Search for new phenomena in events with
  an energetic jet and missing transverse momentum in $pp$ collisions at $\sqrt
  {s}$ =13 TeV with the ATLAS detector}},
  \href{https://doi.org/10.1103/PhysRevD.103.112006}{\emph{Phys. Rev. D}
  {\bfseries 103} (2021) 112006}
  [\href{https://arxiv.org/abs/2102.10874}{{\ttfamily 2102.10874}}].

\bibitem{ATLAS:2020yzc}
{\scshape ATLAS} collaboration, \emph{{Search for dark matter produced in
  association with a single top quark in $\sqrt{s}=13$ TeV $pp$ collisions with
  the ATLAS detector}},
  \href{https://doi.org/10.1140/epjc/s10052-021-09566-y}{\emph{Eur. Phys. J. C}
  {\bfseries 81} (2021) 860}
  [\href{https://arxiv.org/abs/2011.09308}{{\ttfamily 2011.09308}}].

\bibitem{CMS:2020ulv}
{\scshape CMS} collaboration, \emph{{Search for dark matter produced in
  association with a leptonically decaying Z boson in proton-proton collisions
  at $\sqrt{s} =$ 13 TeV}},
  \href{https://doi.org/10.1140/epjc/s10052-020-08739-5}{\emph{Eur. Phys. J. C}
  {\bfseries 81} (2021) 13} [\href{https://arxiv.org/abs/2008.04735}{{\ttfamily
  2008.04735}}].

\bibitem{abdallah2014simplified}
J.~Abdallah, A.~Ashkenazi, A.~Boveia, G.~Busoni, A.~De~Simone, C.~Doglioni
  et~al., \emph{Simplified models for dark matter and missing energy searches
  at the lhc}, {\emph{arXiv preprint arXiv:1409.2893} (2014) }.

\bibitem{Abdallah:2015ter}
J.~Abdallah et~al., \emph{{Simplified Models for Dark Matter Searches at the
  LHC}}, \href{https://doi.org/10.1016/j.dark.2015.08.001}{\emph{Phys. Dark
  Univ.} {\bfseries 9-10} (2015) 8}
  [\href{https://arxiv.org/abs/1506.03116}{{\ttfamily 1506.03116}}].

\bibitem{Csaki:1996ks}
C.~Csaki, \emph{{The Minimal supersymmetric standard model (MSSM)}},
  \href{https://doi.org/10.1142/S021773239600062X}{\emph{Mod. Phys. Lett. A}
  {\bfseries 11} (1996) 599}
  [\href{https://arxiv.org/abs/hep-ph/9606414}{{\ttfamily hep-ph/9606414}}].

\bibitem{Haber:1997if}
H.E.~Haber, \emph{{The Status of the minimal supersymmetric standard model and
  beyond}}, \href{https://doi.org/10.1016/S0920-5632(97)00688-9}{\emph{Nucl.
  Phys. B Proc. Suppl.} {\bfseries 62} (1998) 469}
  [\href{https://arxiv.org/abs/hep-ph/9709450}{{\ttfamily hep-ph/9709450}}].

\bibitem{Dolle:2009fn}
E.M.~Dolle and S.~Su, \emph{{The Inert Dark Matter}},
  \href{https://doi.org/10.1103/PhysRevD.80.055012}{\emph{Phys. Rev. D}
  {\bfseries 80} (2009) 055012}
  [\href{https://arxiv.org/abs/0906.1609}{{\ttfamily 0906.1609}}].

\bibitem{Goudelis:2013uca}
A.~Goudelis, B.~Herrmann and O.~St\r{a}l, \emph{{Dark matter in the Inert
  Doublet Model after the discovery of a Higgs-like boson at the LHC}},
  \href{https://doi.org/10.1007/JHEP09(2013)106}{\emph{JHEP} {\bfseries 09}
  (2013) 106} [\href{https://arxiv.org/abs/1303.3010}{{\ttfamily 1303.3010}}].

\bibitem{georgi1993effective}
H.~Georgi, \emph{Effective field theory}, {\emph{Annual review of nuclear and
  particle science} {\bfseries 43} (1993) 209}.

\bibitem{Burgess:2007pt}
C.P.~Burgess, \emph{{Introduction to Effective Field Theory}},
  \href{https://doi.org/10.1146/annurev.nucl.56.080805.140508}{\emph{Ann. Rev.
  Nucl. Part. Sci.} {\bfseries 57} (2007) 329}
  [\href{https://arxiv.org/abs/hep-th/0701053}{{\ttfamily hep-th/0701053}}].

\bibitem{Backovic:2015soa}
M.~Backovi\'c, M.~Kr\"amer, F.~Maltoni, A.~Martini, K.~Mawatari and M.~Pellen,
  \emph{{Higher-order QCD predictions for dark matter production at the LHC in
  simplified models with s-channel mediators}},
  \href{https://doi.org/10.1140/epjc/s10052-015-3700-6}{\emph{Eur. Phys. J. C}
  {\bfseries 75} (2015) 482}
  [\href{https://arxiv.org/abs/1508.05327}{{\ttfamily 1508.05327}}].

\bibitem{Neubert:2015fka}
M.~Neubert, J.~Wang and C.~Zhang, \emph{{Higher-Order QCD Predictions for Dark
  Matter Production in Mono-$Z$ Searches at the LHC}},
  \href{https://doi.org/10.1007/JHEP02(2016)082}{\emph{JHEP} {\bfseries 02}
  (2016) 082} [\href{https://arxiv.org/abs/1509.05785}{{\ttfamily
  1509.05785}}].

\bibitem{CMS:2018yfx}
{\scshape CMS} collaboration, \emph{{Search for invisible decays of a Higgs
  boson produced through vector boson fusion in proton-proton collisions at
  $\sqrt{s} =$ 13 TeV}},
  \href{https://doi.org/10.1016/j.physletb.2019.04.025}{\emph{Phys. Lett. B}
  {\bfseries 793} (2019) 520}
  [\href{https://arxiv.org/abs/1809.05937}{{\ttfamily 1809.05937}}].

\bibitem{Dutta:2017lny}
B.~Dutta, G.~Palacio, J.D.~Ruiz-Alvarez and D.~Restrepo, \emph{{Vector Boson
  Fusion in the Inert Doublet Model}},
  \href{https://doi.org/10.1103/PhysRevD.97.055045}{\emph{Phys. Rev. D}
  {\bfseries 97} (2018) 055045}
  [\href{https://arxiv.org/abs/1709.09796}{{\ttfamily 1709.09796}}].

\bibitem{Abdallah:2014hon}
J.~Abdallah et~al., \emph{{Simplified Models for Dark Matter and Missing Energy
  Searches at the LHC}}, {\emph{arXiv preprint arXiv:1409.2893} (2014) }
  [\href{https://arxiv.org/abs/1409.2893}{{\ttfamily 1409.2893}}].

\bibitem{Buckley:2014fba}
M.R.~Buckley, D.~Feld and D.~Goncalves, \emph{{Scalar Simplified Models for
  Dark Matter}}, \href{https://doi.org/10.1103/PhysRevD.91.015017}{\emph{Phys.
  Rev. D} {\bfseries 91} (2015) 015017}
  [\href{https://arxiv.org/abs/1410.6497}{{\ttfamily 1410.6497}}].

\bibitem{Degrande:2011ua}
C.~Degrande, C.~Duhr, B.~Fuks, D.~Grellscheid, O.~Mattelaer and T.~Reiter,
  \emph{{UFO - The Universal FeynRules Output}},
  \href{https://doi.org/10.1016/j.cpc.2012.01.022}{\emph{Comput. Phys. Commun.}
  {\bfseries 183} (2012) 1201}
  [\href{https://arxiv.org/abs/1108.2040}{{\ttfamily 1108.2040}}].

\bibitem{Alwall:2014hca}
J.~Alwall, R.~Frederix, S.~Frixione, V.~Hirschi, F.~Maltoni, O.~Mattelaer
  et~al., \emph{{The automated computation of tree-level and next-to-leading
  order differential cross sections, and their matching to parton shower
  simulations}}, \href{https://doi.org/10.1007/JHEP07(2014)079}{\emph{JHEP}
  {\bfseries 07} (2014) 079} [\href{https://arxiv.org/abs/1405.0301}{{\ttfamily
  1405.0301}}].

\bibitem{Sjostrand:2006za}
T.~Sjostrand, S.~Mrenna and P.Z.~Skands, \emph{{PYTHIA 6.4 Physics and
  Manual}}, \href{https://doi.org/10.1088/1126-6708/2006/05/026}{\emph{JHEP}
  {\bfseries 05} (2006) 026}
  [\href{https://arxiv.org/abs/hep-ph/0603175}{{\ttfamily hep-ph/0603175}}].

\bibitem{Sjostrand:2014zea}
T.~Sj\"ostrand, S.~Ask, J.R.~Christiansen, R.~Corke, N.~Desai, P.~Ilten et~al.,
  \emph{{An introduction to PYTHIA 8.2}},
  \href{https://doi.org/10.1016/j.cpc.2015.01.024}{\emph{Comput. Phys. Commun.}
  {\bfseries 191} (2015) 159}
  [\href{https://arxiv.org/abs/1410.3012}{{\ttfamily 1410.3012}}].

\bibitem{deFavereau:2013fsa}
{\scshape DELPHES 3} collaboration, \emph{{DELPHES 3, A modular framework for
  fast simulation of a generic collider experiment}},
  \href{https://doi.org/10.1007/JHEP02(2014)057}{\emph{JHEP} {\bfseries 02}
  (2014) 057} [\href{https://arxiv.org/abs/1307.6346}{{\ttfamily 1307.6346}}].

\bibitem{Florez:2019tqr}
A.~Fl\'orez, A.~Gurrola, W.~Johns, J.~Maruri, P.~Sheldon, K.~Sinha et~al.,
  \emph{{Anapole Dark Matter via Vector Boson Fusion Processes at the LHC}},
  \href{https://doi.org/10.1103/PhysRevD.100.016017}{\emph{Phys. Rev. D}
  {\bfseries 100} (2019) 016017}
  [\href{https://arxiv.org/abs/1902.01488}{{\ttfamily 1902.01488}}].

\end{thebibliography}\endgroup







\end{document}